\documentclass[twocolumn,showpacs,amsmath,floatfix,prb,aps]{revtex4}
\usepackage{graphicx}  % needed for figures
\usepackage{dcolumn}   % needed for some tables
\usepackage{bm}        % for math
\usepackage{amssymb}   % for math
\usepackage[usenames]{color}
\usepackage{url}
\usepackage{hyperref}
\newcommand{\tokyo}{Department of Applied Physics, The University of Tokyo, Tokyo 113-8656, Japan}
\begin{document}

%%\title{\textcolor{red}{\textit{Ab-initio}} calculations on \textcolor{red}{intrinsic} interface states \textcolor{red}{controlleld by atomic stacking sequence} at SiC/SiO$_2$ interface}
\title{A novel intrinsic interface state controlled by atomic stacking sequence at interfaces of SiC/SiO$_2$}

\author{Yu-ichiro Matsushita}   \affiliation{\tokyo}
\author{Atsushi Oshiyama}         \affiliation{\tokyo}
\date{\today}

\begin{abstract}
On the basis of \textit{ab-initio} total-energy electronic-structure calculations, we find that interface localized electron states at the SiC/SiO$_2$ interface emerge in the energy region between 0.3 eV below and 1.2 eV above the bulk conduction-band minimum (CBM) of SiC, being sensitive to the sequence of atomic bilayers in SiC near the interface. These new interface states unrecognized in the past are due to the peculiar characteristics of the CBM states which are distributed along the crystallographic channels. We also find that the electron doping modifies the energetics among the different stacking structures. Implication for performance of electron devices fabricated on different SiC surfaces are discussed. 
\end{abstract}

\pacs{~}
\maketitle

%\section{Introduction}
High-efficiency power electronic devices play an important role in realization of the energy-saving society. To increase the efficiencies of the power devices, low energy-loss semiconductor materials are necessary. SiC has attracted much attention as a possible next-generation power semiconductor due to its prominent material properties such as high breakdown electric field (10 times larger than Si) and high thermal conductance (3 times larger than Si) \cite{Kimoto,Baliga}. Another benefit of SiC power semiconductor is the utility of its 
natively oxidized thin films, SiO$_2$, for the fabrication of metal-oxide-semiconductor field-effect transistors (MOSFETs), assuring the good connectivity with Si technology \cite{Kimoto}.

SiC-MOSFET devices have been already available commercially. However, they still face a severe problem that the mobility of the devices is far from the theoretical values due to the huge density of interface levels at the SiC/SiO$_2$ with the concentration of 10$^{13}$ - 10$^{14}$ cm$^{-2}$ eV$^{-1}$ \cite{Kimoto,Yoshioka,Kobayashi}. The levels appearing in the gap within the range of 0.3 eV below the CBM indeed cause the substantial reduction of the electron mobility \cite{Kimoto}. Many theoretical and experimental efforts have been done to identify those interface levels and carbon-related defects are suspicious of the mobility killers \cite{Afanasev,Kikuchi,Gali1,Gali2,Gali3,Kobayashi2}. However, no consensus is reached yet. In this Letter, we show that, neither defects nor impurities, but the imperfection in the stacking sequence of the atomic layers causes interface levels below the CBM. 

SiC is a tetrahedrally bonded semiconductor in which atomic bilayers consisting of Si and C atoms are stacked along the bond direction. Different stacking sequences lead to different crystal structures called polytypes and each structure is labeled by the stacking sequence: The most frequently obtained structure is 4H-SiC whose stacking sequence is ABCB of 4-bilayer periodicity with hexagonal symmetry. Although the local atomic structures of the polytypes are idential to each other, their electronic properties, in particular the band gaps, are known to differ from one to another \cite{Kimoto}. As we have clarified in Ref.~\onlinecite{Matsushita1,Matsushita2,Matsushita3}, this is due to the surprisingly interesting character of the conduction band minimum (CBM): i.e., the wavefunction of the CBM is not distributed around the atomic sites but extended or floats in the interstitial channels generally existing in the tetrahedrally bonded structures \cite{Matsushita4}. This \textit{floating} nature renders the energy level of the CBM being strongly affected by the length of the internal channel which is peculiar to each polytype.

From moment to moment during the thermal oxidation of 4H-SiC (0001) surface, only two types of the stacking termination at the interface are possible in case of the layer-by-layer oxidation: One is a cubic interface (BCBA-stack/SiO$_2$), and the other is a hexagonal interface (ABCB-stack/SiO$_2$). In addition to those interfaces, we here consider the stacking fault at the interface. Actually, it is experimentally reported that the stacking sequences are transformed at the SiC surface, i.e., ABCA-stacking order \cite{Starke}. The similar stacking imperfection is likely to occur also at the interface, leading to the stacking-fault interface (ABCA-stack/SiO$_2$). As deduced from the \textit{floating} nature explained above, the variation of the stacking sequence near the interface leads to the variation of the channel length there and hereby varies the energy level of the CBM at the interface. In this Letter, based on the density-functional theory (DFT) \cite{DFT,HK}, we find that the stacking-fault interface induces a level in the gap at 0.3 eV below the CBM, thus being a strong and intrinsic candidate for the mobility killer. We also find that such stacking-fault interface structure is energetically favorable in the negatively charged interface. 

%\section{Calculation details}
Full geometry optimization for all systems was performed using the Vienna \textit{ab-initio} Simulation Package (VASP) with PAW pseudo potential method~\cite{VASP1,VASP2} using PBE exchange-correlation potential~\cite{GGA-PBE} in the generalized-gradient approximation(GGA). In this study, we have considered the (0001) surface and adopted a SiC slab model consisting of 8 bilayers with the $\sqrt{3} \times \sqrt{3}$ periodicity in the lateral plane. A vacuum region of 20-\AA\ thickness is enough to avoid fictitious interactions between the adjacent slabs. The bottom surface atoms of the slab is fixed to the bulk crystallographic positions and terminated by H atoms to compensate for the missing bonds, whereas the rest of the system is allowed to evolve and relax freely. The energy cutoff of 400\,eV and a $\Gamma$-centered Monkhorst-Pack $5\times 5\times 1$ $k$-point grid were used. These parameters are adopted after the examination of the accuracy within 8 meV per atom in total energy. The structural optimization has been done with a tolerance of $10^{-1}$ eV\AA$^{-1}$.

%\section{Results}

\begin{figure}
\includegraphics[width=1.0\linewidth]{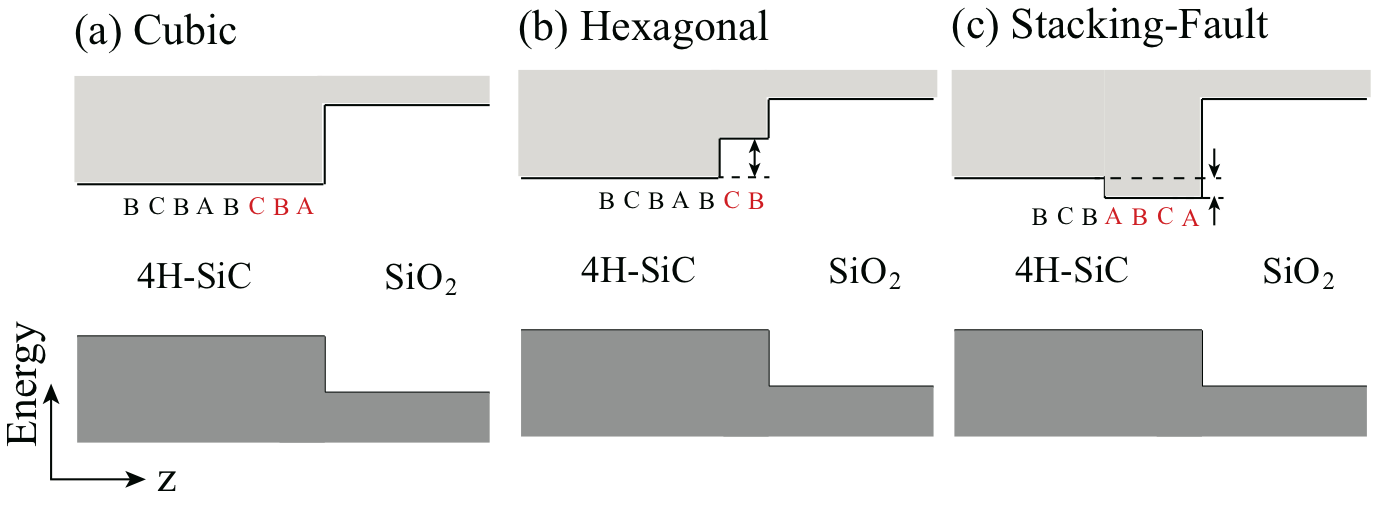}
\caption{
(Color online). 
Schematic pictures of the local density of states (LDOS) along the direction perpendicular to the interface (z-direction) for the three possible interface structures of SiC/SiO$_2$: (a) The cubic interface (BCBA-stack/SiO$_2$), (b) the hexagonal interface (ABCB-stack/SiO$_2$), and (c) the stacking-fault interface (ABCA-stack/SiO$_2$). The stacking sequence near the interface is shown by the letters. The red letters denote the region in which the interstitial channel is connected.
}
\label{Fig1}
\end{figure}

\begin{table}
 \label{Energy_comparison}
\caption{
Calculated total energies of the non-doped and electron-doped SiC (0001) surfaces with the three different bilayer atomic sequences with respect to the most stable structure in unit of meV per 54-atom unit cell.
}
\begin{tabular}{l|c|c}
\hline
Surface Stacking & \ non-doped \ & electron-doped \\
\hline 
cubic (BCBA/)          & 0  & 87 \\ 
hexagonal (ABCB/ )     & 33 & 196 \\ 
stacking-fault (ABCA/) & 48 & 0 \\ 
\hline 
\end{tabular}
\end{table}

We have investigated the three interfaces with different stacking sequences near the interface: i.e., the cubic (BCBA-stack/SiO$_2$), the hexagonal (ABCB-stack/SiO$_2$) and the stacking-fault (ABCA-stack/SiO$_2$) interfaces, as shown in Fig.~\ref{Fig1}. It is noteworthy that the channel lengths near the interface in the cubic, the hexagonal and the stacking-fault interfaces are 3, 2, and 4, respectively, in unit of the bilayer thickness. When the wavefunction is confined in a shorter channel space, the corresponding energy level of the CBM is expected to shift upward because of the quantum confinement. This leads to the distinct energy diagrams for the three interface structures, schematically shown in Fig.~\ref{Fig1} and also quantitatively revealed below in Figs.~\ref{Fig2} and \ref{Fig4}. In particular, the interface state appears below the CBM in the stacking-fault interface.

To validate our argument above, we start with the energetics among the structures with different bilayer stacking. For the purpose, we consider the cubic, the hexagonal and the stacking-fault SiC slabs in which the topmost atomic layer is terminated by H atoms to mimic the SiO$_2$ layers. The calculated total energies are shown in the second column of Table.~I. The cubic sequence is the lowest, the hexagonal the second lowest, and the stacking-fault the highest. However, the total energy difference is small, less than 1 meV per atom, indicating that the stacking imperfection is likely to occur in real situations.

\begin{figure}
\includegraphics[width=0.7\linewidth]{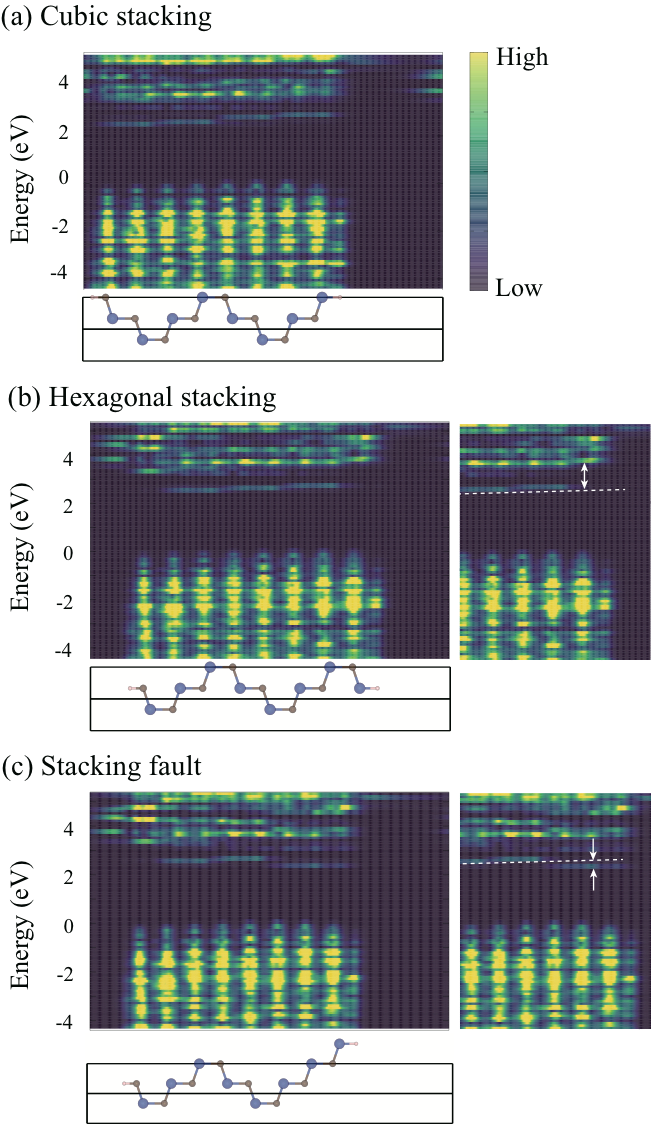}
\caption{(Color online). 
Calculated LDOS for the cubic-stacking (BCBA-stack) (a), the hexagonal-stacking (ABCB-stack) (b), and the stacking-fault (ABCA-stack) (c) SiC surfaces as a function of the energy and the z-coordinate perpendicular to the surface. The magnitude of LDOS is presented by the color code shown in the legend. The left and right sides correspond to the bottom and top surfaces, respectively, of the SiC slab. The origin of the energy is set to be the valence band top at the top-surface region. The right panels in (b) and (c) are enlarged LDOS near the top surface and the dashed line is the guide for eyes to make the position of the bulk CBM clearly (see text). Below each LDOS, corresponding atomic configuration is illustrated where blue, brown and white balls depict Si, C and H atoms, respectively.
}
\label{Fig2}
\end{figure}

Characteristics of electron states near surfaces or interfaces manifest themselves in the local density of states (LDOS) which is defined as
\begin{eqnarray}
{\rm LDOS}(\epsilon,z)=\int d{\bf r}_\perp \sum_{n {\bf k}}\delta(\epsilon-\epsilon_{n{\bf k}})|\phi_{n{\bf k}}({\bf r})|^2,
\end{eqnarray}
where $d{\bf r}_\perp$ represents a two-dimensional vector in the plane parallel to the surface or the interface, $\epsilon$ is an energy, $\phi_{n{\bf k}}({\bf r})$ is a wavefunction (Kohn-Sham orbital in DFT), and $\epsilon_{n{\bf k}}$ is an eigenvalue. Figure~\ref{Fig2} shows calculated LDOS for the three different surfaces, i.e., the cubic and the hexagonal stacking and the stacking-fault surfaces of SiC(0001). In the valence band region (negative energy region in Fig.~\ref{Fig2}), we observe spiky spectra representing  
the eigenstates of the valence bands. Their positions in real space correspond to the positions of the atomic layers. This means that valence electrons are distributed around atoms. On the other hand, the conduction bands have no such spiky structures in LDOS. This reflects the fact that the conduction electrons are distributed in internal channel space broadly. The calculated energy gap represented by the dark blue region in the slab in Fig.~\ref{Fig2} is 2.3 eV, smaller than the experimental value of 3.3 eV for 4H-SiC, in our GGA calculations.

First, we notice
that the band lineup along the direction perpendicular to the surface is slanted. 
This is due to 
the internal dielectric polarization induced by the low symmetry of the 4H-SiC. The dashed lines in the right panels 
in Fig.~\ref{Fig2} represent the slanted band lineup caused by such internal electric field. 
Second, 
comparing the Figs.~\ref{Fig2}(a) and (b), we have found that the hexagonal-stacking (ABCB/) surface induces a surface state located about 1.2 eV above the bulk CBM. This interface state is caused by the quantum confinement of the bulk CBM state in the interstitial channel near the surface: The length of the channel is shorter near the hexagonal surface as stated above. More importantly, in the stacking-fault (ABCA/) surface, the interface state caused by the modulation of the length of the interstitial channel is located at 0.3 eV below the bulk CBM [Fig.~\ref{Fig2}(c)]. This unequivocally clarifies that the stacking difference near the surface changes the surface properties considerably. 

Band bending near the surface or the interface as well as the polarity of SiC may cause electron doping in the conduction-band states in Fig.~\ref{Fig2}. This may change the energetics among the different stacking-sequence structures. 
We have actually performed calculations for the electron-doped surfaces with the concentration of $4\times 10^{14}$ cm$^{-2}$. In our calculations, the doped electrons occupy the surface conduction states and 
modifies the energetics (See the third column of Table~I). 
The surface structure with the stacking-fault (ABCA/) is lower in total energy than 
those 
of the cubic (BCBA/) and the hexagonal (ABCB/) surfaces by 0.1 eV and 0.2 eV, respectively.

\begin{figure}
\includegraphics[width=0.6\linewidth]{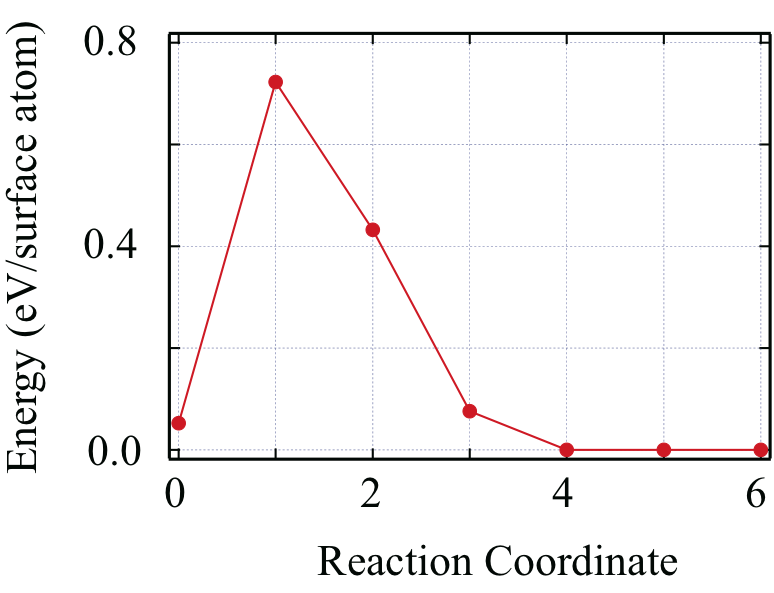}
\caption{(Color online). Calculated energy barrier of the stacking transformation from the hexagonal-stacking (ABCB/) to the stacking-fault (ABCA/) structures on SiC(0001) surface.
}
\label{Fig3}
\end{figure}

We have clarified above that the new surface states appear near the CBM in the region between CBM $-$ 0.3 eV (the case of the stacking-fault surface) $\sim$ CBM + 1.0 eV (the case of the hexagonal surface), depending on the atomic-bilayer stacking near the surface. This stacking-dependent appearance of the electron state is found also in the SiC/SiO$_2$ interface (see below). We have also clarified that the structures with different stacking sequences have comparable total energies. Then we have next 
calculated the energy barrier of the 
transformation between the two different stacking sequences, i.e, 
from the hexagonal stacking structure (ABCB/) to the stacking-fault structure (ABCA/) 
for the electron-doped system, 
using the nudged-elastic-band (NEB) method\cite{NEB}.  
Through the transformation, each atom of the topmost layer moves only 1.9 {\AA} in the top atomic plane. 
The calculated energy for the transformation is shown in Fig.~\ref{Fig3}.  As stated above, 
the final stacking-fault structure is more stable than the initial hexagonal-stacking structure, showing an exothermic reaction pathway. 
We have further clarified 
that the energy barrier for this stacking transformation is 0.8 eV per surface atom. 
We have also calculated the energy barriers for the stacking transformations without electron doping and found that they are about 1 eV per atom.

\begin{figure}
\includegraphics[width=0.7\linewidth]{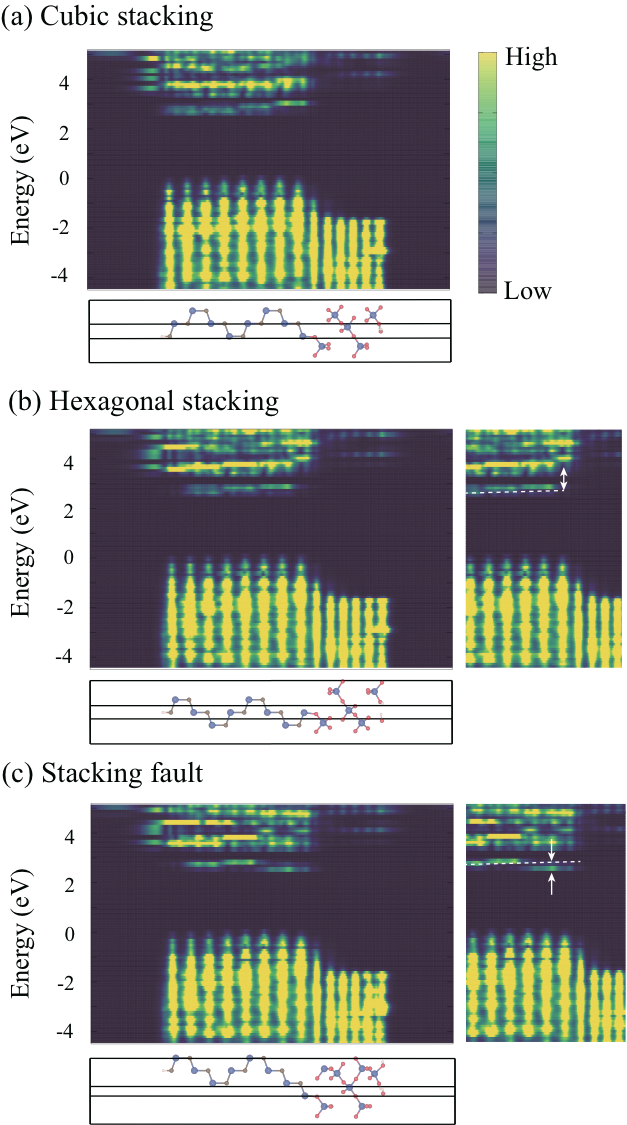}
\caption{(Color online). 
Calculated LDOS for the cubic-stacking (BCBA/SiO$_2$) (a), the hexagonal-stacking (ABCB/SiO$_2$) (b), and the stacking-fault (ABCA/SiO$_2$) (c) SiC/SiO$_2$ interfaces as a function of the energy and the z-coordinate perpendicular to the interface. The magnitude of LDOS is presented by the color code shown in the legend. The left and right sides correspond to the SiC and SiO$_2$ regions. The origin of the energy is set to be the valence band top at the top-surface region of SiC. The right panels in (b) and (c) are enlarged LDOS near the interface and the dashed line is the guide for eyes to make the position of the bulk CBM of SiC clearly (see text). Below each LDOS, corresponding atomic configuration is illustrated where blue, brown, red and white balls depicts Si, C , O and H atoms, respectively.
}
\label{Fig4}
\end{figure}

The substantial modification of the electron states near the CBM of SiC due to the stacking difference of atomic bilayers has been also found in SiC/SiO$_2$ interface by our GGA calculations.
We take the most stable crystalline form of SiO$_2$, $\alpha$-quartz, to model real amorphous SiO$_2$, and have optimized the structures for the interface structures with the three different stacking sequences: i.e., the cubic-stacking (BCBA/SiO$_2$), the hexagonal-stacking (ABCB/SiO$_2$), and the stacking-fault (ABCA/SiO$_2$). The Si dangling bonds emerged at the interface have been terminated by H atoms. 
Fig.~\ref{Fig4} shows the calculated LDOS 
for the three interface structures. Our calculations show that the valence-band offset is 1.5 eV common to the three interface structures. 
In contrast, the electron states are sensitive near the CBM to the bilayer stacking sequence: There emerges the new electron state which is distributed at the interface and is located at 0.3 eV below (the stacking-fault interface) and 1.0 eV above (the hexagonal stacking interface) the bulk CBM. 
Correspondingly, the band offset of the CBM takes values from 0.6 to 1.6 eV. 
We emphasize that this variation of the electronic structure and even the appearance of the interface levels in the gap are due to the \textit{floating} nature of the CBM state of SiC. 

Imperfection of atomic stacking is commonly observed in tetrahedrally bonded semiconductors. This planar imperfection has been thought to play a minor role in electronic structure. However, We have found, for the SiC(0001) surface or the interface, that this stacking sequence determines the length of the internal channel and thus induces an interface state which is crucial in the performance of MOSFET devices. In SiC MOSFET, non-polar surfaces such as $(11\bar{2}0)$- or $(1\bar{1}00)$-face are occasionally 
used for the device fabrication. In those non-polar surfaces, 
the lengths of the cannels are infinite, thus being independent of the bilayer stacking along the (0001) direction.
Hence, the interface state near the CBM is not expected to emerge by the stacking modulation on the non-polar surface. From this viewpoint, the $(11\bar{2}0)$ and $(1\bar{1}00)$ surfaces are expected to have advantages than the (0001) surface to fabricate high-performance SiC devices.

%\section{Summary}
To summarize, 
on the basis of the density-functional calculations, we have elucidated that the imperfection of the atomic-stacking sequence near the SiC/SiO$_2$ interface induces interface levels at 0.3 eV below the conduction band bottom of SiC, thereby proposing the stacking fault as an intrinsic killer of the carrier mobility. We have also shown that the stacking-fault interface structure has comparable total energy with the perfect-stacking structures and even becomes the most stable upon electron doping. Underlying physics of all these findings is the \textit{floating} nature of the conduction-band state of SiC.

\begin{acknowledgments}
We thank discussion with Professor Kenji Shiraishi. Computations were performed mainly at the Center for Computational Science, University of Tsukuba, and the Supercomputer Center at the Institute for Solid State Physics, The University of Tokyo. Y.M. acknowledges the support from JSPS Grant-in-Aid for Young Scientists (B) (Grant Number 16K18075).
\end{acknowledgments}

%\bibliography{tet_paper}
\end{document}